\documentclass[12pt,a4paper]{iopart}
\usepackage[german,english]{babel}
\usepackage[latin1]{inputenc}
\usepackage{amssymb}
\usepackage{cite}
\usepackage{bbm}
\usepackage{graphicx}

\newcommand{\dd}{\mathrm{d}}
\newcommand{\one}[1]{\mathop{#1}\limits^1}
\newcommand{\two}[1]{\mathop{#1}\limits^2}%
\begin{document}

\title[Impurity in the SUSY $t$--$J$ model]{Interplay between a quantum
  impurity and a boundary field in the SUSY $t$--$J$ model}

\author{Holger Frahm and Guillaume Palacios}

\address{Institut f{\"ur} Theoretische Physik, Leibniz Universit{\"a}t
  Hannover,\newline Appelstr.\,2, 30167~Hannover, Germany}

\begin{abstract}
We study the role of bound states appearing in different formulations of the
Bethe ansatz for the supersymmetric $t$--$J$ model with a boundary potential
and an integrable impurity.  For special values of the parameters describing
the boundary and the impurity the charge fluctuations at the latter vanish.
The population of the bound states selects different sectors of the impurity
levels leading to integrable Kondo impurities.  
%
\end{abstract}

\maketitle

\section{Introduction}  %
\label{sec:intro}
The influence of quantum impurities embedded into a system of conducting
electrons on the properties of the host have long been the subject of research
activities in condensed matter physics.  Theoretically, local scatterers with
internal degrees of freedom in a lattice system can be described by the
Anderson and Kondo models \cite{Hewson}.
In recent years these models have found new applications beyond their original
realm of magnetic impurities in a metallic system: electron transport through
quantum dots and wires or molecules in contact to a metallic surface provide
realizations of the Kondo effect which allows for experimental control of the
relevant parameters.  As a consequence, in particular the behaviour of a Kondo
impurity embedded into a Luttinger liquid has been investigated in great
detail using field theoretical methods \cite{LeTo92,FuNa94,FrJo95,EgKo98}.
With these methods the critical behaviour of such systems can be classified.
Still for the full picture and in particular for a better understanding of the
emergence of this behaviour within a microscopic realization exact solutions
of integrable lattice models provide useful insights.

Following the Bethe Ansatz solution of the Kondo Hamiltonian
\cite{TsWi83,AnFL83} several one-dimensional models dealing with impurities
have been constructed within the framework of the quantum inverse scattering
method, e.g.\ the Heisenberg chain with an magnetic impurity \cite{AnJo84} or
integrable inhomogeneities in a $t$--$J$ chain
\cite{BEF96,BEF97,ScZv97,BoKS05}.  Within this approach the hybridization of
the local level with the host system can be varied without spoiling
integrability while the inhomogeneity's degrees of freedom have to be
compatible with the symmetry of the host system.  Additional control of the
impurity is possible by combining it with a physical boundary
\cite{FrZv97b,BeFr99a}.  Here the strength of a scalar boundary potential
appears as an additional parameter and certain restrictions on the spectrum of
the local scatterer can be relaxed.  This allows for the construction of
purely magnetic (Kondo) impurities in one-dimensional lattice models of
correlated itinerant electrons \cite{WDHP97,ZhGo99,ZGLG99}.  Within Sklyanin's
reflection algebra \cite{Cher84,Skly88} these new impurity models arise as a
consequence of a synchronization of the boundary potential to the parameters
of the impurity where different sectors in the impurity's internal Hilbert
space decouple \cite{FrSl99} allowing to project out some of the local
configurations -- similar as in the Schrieffer-Wolff transformation from the
Anderson to the Kondo model.
This projection (performed on the level of the Hamiltonian) has a direct
influence on the determination of the many-particle spectrum by means of the
algebraic Bethe ansatz: to capture the influence of the impurity on the
physical properties it has to be ensured that the correct sector of the
impurity's state space is kept.  This can be achieved by working directly
with the projected system \cite{ZhGo99,ZGLG99}.  Alternatively, it should be
possible to extract the levels from the full impurity spectrum provided that a
criterion can be formulated which allows to select states from a given sector.

In this paper we show that this selection is indeed possible based on the
presence or absence of special solutions -- identified with bound states in
the spectrum -- to the Bethe equations which are known to exist in either the
presence of a sufficiently strong boundary field or that of an impurity.  We
begin with a brief review of the construction of integrable lattice models
with boundaries.  For a particular inhomogeneity in the supersymmetric
$t$--$J$ model already considered in \cite{FrSl99} we analyze the spectrum of
impurity and boundary bound states and then show how different sets of Bethe
ansatz equations for a Kondo spin-$s$ emerge from the impurity system when
these bound states are populated.

\section{Integrable impurities combined with boundaries}  
\label{sec:qism}
To introduce our notations we start with a brief review of the theory of
integrable models with boundaries. In the framework of the quantum inverse
scattering method (QISM) \cite{VladB} the construction of integrable
Hamiltonians is based on vertex models obtained by combining ${\cal
L}$-operators which satisfy the intertwining relation
\begin{equation}
  {\cal R}^{12}(\lambda-\mu)
        \left({\cal L}(\lambda)\otimes{\cal L}(\mu)\right)=
  \left({\cal L}(\mu)\otimes{\cal L}(\lambda)\right)
        {\cal R}^{12}(\lambda-\mu)\ .
\label{IR}
\end{equation}
Models which are constructed in this approach are classified by a particular
choice of the ${\cal R}$-matrix entering (\ref{IR}) which in turn has to solve
the quantum Yang-Baxter equation (YBE):
\begin{equation}
  {\cal R}^{12}(\lambda){\cal R}^{13}(\lambda+\mu){\cal R}^{23}(\mu)=
  {\cal R}^{23}(\mu){\cal R}^{13}(\lambda+\mu){\cal R}^{12}(\lambda)\,.
\label{YBE}
\end{equation}
(Superscripts denote the spaces in the tensor product $V_1\otimes V_2\otimes
V_3$ in which ${\cal R}^{ij}$ acts nontrivially).  Of particular interest are
models constructed based on the `fundamental' rational ${\cal R}$-matrices
${\cal R}^{ij}(\lambda) = (\lambda \mathbf{1} +i\Pi^{ij})/(\lambda+i)$ where
$\Pi^{ij}$ is a permutation operator on the space $V_i\otimes V_j$.  Different
representations ${\cal L}$ of the quadratic algebra (\ref{IR}) for a given 
${\cal R}$  can be combined to construct integrable lattice models.
For example, choosing ${\cal L}_0^n(\lambda)={\cal R}^{0n}(\lambda)$ and
identifying $V_n$ with the quantum space corresponding to site $n=1,\ldots,L$
of a one-dimensional lattice a family of commuting operators on the space
$V_1\otimes\ldots\otimes V_L$ is given by the transfer matrix (products of the
${\cal L}$-operators and the trace are taken in the `auxiliary' space $V_0$)
\begin{equation}
  t_L(\lambda) = \mathrm{tr}_0\left({\cal L}_0^L(\lambda){\cal
  L}_0^{L-1}(\lambda)\ldots {\cal L}_0^1(\lambda)\right)\ .
\end{equation}
The fundamental rational models with nearest neighbour interaction obtained
within this approach are, among others, the spin $S=\frac{1}{2}$ Heisenberg
chain and the one-dimensional supersymmetric $t$--$J$ model with periodic
boundary conditions.

Integrable inhomogeneities can be inserted into such a system by replacing the
${\cal L}$-operator at one site of the lattice by a different solution ${\cal
L}_{\mathrm{imp}}$ of the intertwining relation acting on $V_0\otimes
V_\mathrm{imp}$ (see e.g.\ Refs.~\cite{AnJo84,BEF96,BEF97,ScZv97}).
In ${\cal L}_{\mathrm{imp}}$ the internal quantum degrees of freedom of the
inhomogeneity are controlled by the choice of a specific representation of the
underlying algebra acting on the quantum space $V_\mathrm{imp}$ which may be
different from the one used for the other (bulk) sites (i.e.\ $SU(2)$ for the
Heisenberg chain, $gl(2|1)$ for the supersymmetric $t$--$J$ model).  In
addition the coupling of the imhomogeneity site to the rest of the lattice can
be varied by a shift of the argument, i.e.\ ${\cal
L}_{\mathrm{imp}}(\lambda)\to{\cal L}_{\mathrm{imp}}(\lambda+t)$ which is
consistent with relation (\ref{IR}).

In this paper we consider an inhomogeneity in the supersymmetric $t$--$J$
model acting on $V_\mathrm{imp}$ through an `atypical' representation $[s]_+$
of $gl(2|1)$ (see Refs.~\cite{snr:77,Marcu80} and \ref{app:gl21}). This
particluar representation has already been used in constructing an integrable
model of a doped spin $s$ Heisenberg chain \cite{FrPT98,Frahm99,FrSl00}.  In
terms of the generators of $gl(2|1)$ the impurity ${\cal L}$-operator is given
by \cite{Kulish85}
\begin{equation}
{\cal L}_\mathrm{imp} (\lambda) \equiv
{\cal L}_s (\lambda)=\frac{\lambda - i(s+1/2)}{\lambda +
i(s+1/2)}\mathbbm{1}+\frac{i}{\lambda +i(s+1/2)}C_2^{(0s)}\,.
\label{LS+}
\end{equation}
Here $C_2^{(0s)}$ is the quadratic Casimir (\ref{cas2}) of $gl(2|1)$ on the
tensor product $V_0\otimes V_\mathrm{imp}$ (the auxiliary space $V_0$ carries
the three-dimensional fundamental representation $[1/2]_+$ of $gl(2|1)$).

Boundary conditions different from periodic ones can be treated within the
QISM by extending the algebra defined by the intertwining relations through
so-called reflection equations (RE) \cite{Cher84,Skly88}.  The RE define two
algebras ${\cal T}_\pm$ whose representations allow for a classification of
integrable boundary conditions.  ${\cal T}_+$ and ${\cal T}_-$ are related by
an algebra automorphism, for ${\cal T}_-$ the RE reads:
\begin{equation}
\fl
  {\cal R}^{12}(\lambda-\mu)
      \one{{\cal T}_-}(\lambda){\cal R}^{21}(\lambda+\mu)
        \two{{\cal T}_-}(\mu)
 =\two{{\cal T}_-}(\mu){\cal R}^{12}(\lambda+\mu)
      \one{{\cal T}_-}(\lambda){\cal R}^{21}(\lambda-\mu)\, ,
\label{REF}
\end{equation}
where ${\cal R}^{12}$ is again the solution of the YBE (\ref{YBE}) on the
tensor product $V_1\otimes V_2$ and $\one{{\cal T}_-}= {\cal T} \otimes I$,
$\two{{\cal T}_-}= I \otimes {\cal T}$.  The representations of ${\cal T}_\pm$
determine the boundary terms in the Hamiltonian at the left (right) end of the
chain.  Since these can be chosen independently it is sufficient to consider
solutions of (\ref{REF}) to obtain a classification of the possible boundary
impurities.
In the following we shall treat the explicit case of an open supersymmetric
$t$--$J$ chain with a boundary chemical potential \cite{Essl96}. 
For this model the possible boundary conditions are determined by $c$-number
solutions $K_\pm(\lambda)$ of the RE (\ref{REF}).  For the fundamental
$gl(2|1)$-symmetric ${\cal R}$-matrix these solutions have been classified
(\cite{Gonz94}):  the present case of a boundary chemical potential term is
described by $K_-^p (\lambda)= \mathrm{diag}\left(1,1,-\frac{p\lambda
+i}{p\lambda -i}\right)$.

Given these solutions to (\ref{IR}) and (\ref{REF}), the commuting integrals
of motion of the system are given by the transfer matrix
\begin{eqnarray}
  \tau_L(\lambda) = && \mathrm{tr}_0
  \Bigl( K_+^{p_L}(\lambda) {\cal L}_0^L(\lambda){\cal
  L}_0^{L-1}(\lambda)\ldots {\cal L}_0^1(\lambda)\, \times
\nonumber\\
  &&\qquad  \times\, K_-^{p_1}(\lambda)
  \left({\cal L}_0^1(-\lambda)\right)^{-1} 
  \left({\cal L}_0^2(-\lambda)\right)^{-1} \ldots 
  \left({\cal L}_0^L(-\lambda)\right)^{-1}
  \Bigr)\ . 
\label{topen}
\end{eqnarray}
In particular, the Hamiltonian is obtained by taking the derivative of
$\tau_L$ at the `shift point' $\lambda=0$.  Choosing the left end of the
$t$--$J$ chain to be purely purely reflecting (i.e.\ $p_1=0$ or
$K_+\equiv\mathbf{1}$) and $K_-^p (\lambda)$ for the right one the result is
given in terms of the bulk contribution (the operator ${\cal P}$ projects out
states with double occupancy on any site of the lattice)
\begin{eqnarray}
H_{tJ} =& -{\cal P}\left( \sum_{j=1}^{L-1}\sum_\sigma c^\dagger_{j,\sigma}
c_{j+1,\sigma} + c^\dagger_{j+1,\sigma} c_{j,\sigma}\right){\cal P}
\nonumber\\
&+2\sum_{j=1}^{L-1}\left[\vec{S}_j \vec{S}_{j+1} -\frac{n_j n_{j+1}}{4} +\frac{1}{2}(n_j + n_{j+1})\right]
-B S^z -\mu N
\label{HtJ}
\end{eqnarray}
where we have added a magnetic field $B$ and a chemical potential $\mu$
coupling to the total magnetization and particle number, respectively.  In
addition one obtaines a boundary contribution containing the coupling to the
boundary potential $p$,
\begin{equation}
H_p = pn_1.
\label{Hp}
\end{equation}

Just as for the case of periodic boundary conditions one can insert integrable
inhomogeneities into the system by adding an additional site described by
${\cal L}_\mathrm{imp}(\lambda+t)$.  Particularly interesting is the case
where this inhomogeneity is placed at one of the boundaries (see e.g.\
\cite{FrZv97b,FrSl99}).  This can be realized by replacing the boundary matrix
$K_-^p$ in Eq.~(\ref{topen}) by an operator valued `dressed' one, i.e.
\begin{equation}
  {\cal K}_-(\lambda)={\cal L}_\mathrm{imp}(\lambda+t)K_-^p(\lambda)
    \left({\cal L}_\mathrm{imp}(-\lambda+t)\right)^{-1}\,.
\label{kdress}
\end{equation}
Here, in addition to the choice of a representation for the impurity ${\cal
L}$-operator and the shift $t$ in the spectral parameter, we can tune the
boundary parameter $p$ to control the properties of the inhomogeneity.

Using the boundary matrix (\ref{kdress}) the boundary contribution to the
Hamiltonian with an $[s]_+$ impurity (\ref{LS+}) is obtained to be
\begin{eqnarray}
\fl
   H_\mathrm{bimp} =& -2p \left(B_1-\frac{1}{2}\right)
      +\frac{1}{t^2 +(s+1/2)^2}\left((2s+1)\mathbbm{1}-C_2^{(s1)}\right)+
\nonumber\\
\fl &       - \frac{2p}{t^2 +(s+1/2)^2}\left(
         it[B_1,C_2^{(s1)}] 
       - (s+\frac{1}{2})\{B_1,C_2^{(s1)}\}
       + C_2^{(s1)}B_1C_2^{(s1)}
     \right)
\label{Hbimp}
\end{eqnarray}
where $B_1 =1-{1 \over 2}n_1$, $C_2^{(s1)}$ is the quadratic Casimir operator
(\ref{cas2}) of $gl(2|1)$ on the quantum space $V_\mathrm{imp}\otimes V_1$ and
$[.,.]$ ($\{.,.\}$) denote (anti-)commutators.  Note that the Hamiltonian is
hermitean for real boundary potentials $p$ and any real $t$ while there are
non-hermitean terms for finite $p$ and imaginary $t$.


\section{Spectrum and bound states} %
\label{sec:bbs}
Both the boundary potential $p$ and the presence of an impurity affect the
nature of the spectrum of the chain.  It is natural to expect that, for
sufficiently strong $p$, boundary bound states (or anti-bound states) are
formed at the end of the chain.  This issue has been studied in the context of
the X-ray edge singularity problem for one-dimensional lattice models of
correlated electrons \cite{EsFr97}.  Similarly, an inhomogeneity can lead to
the formation of bound states as its coupling to the bulk of the system is
varied \cite{FrPa06}.  For the integrable model considered here, both
scenarios can be discussed by the analysis of the Bethe ansatz equations
(BAE).  Let us consider the case of a repulsive boundary potential, $p>0$.
Starting from the fully polarized state which maximizes the number of
particles (the Sutherland pseudo-vacuum \cite{suth:75}) the wave function of
an eigenstate with $N_h$ holes and $N_\downarrow$ overturned spins is
parametrized by the roots $\{\lambda_k\}$ and $\{\vartheta_\ell\}$ of the BAE
\cite{Essl96}
\begin{eqnarray}
\fl
 [e_1(\lambda_k)]^{2L}\eta_{\mathrm{imp}}(\lambda_k)\eta_p (\lambda_k) =
  \prod_{j\ne k}^{N_h
    +N_\downarrow}e_2(\lambda_k-\lambda_j)e_2(\lambda_k+\lambda_j)
\prod_{\ell=1}^{N_h} e_{-1}(\lambda_k-\vartheta_\ell)
e_{-1}(\lambda_k+\vartheta_\ell)\,,
\nonumber\\[-8pt]
\label{Sup}
\\[-8pt]
  1= \xi_{\mathrm{imp}}(\vartheta_\ell)\xi_p (\vartheta_\ell)\prod_{j=1}^{N_h
    +N_\downarrow}e_1(\vartheta_\ell -\lambda_j) e_1(\vartheta_\ell
  +\lambda_j)\,.
\nonumber
\end{eqnarray}
Here we have introduced the function $e_y (x)=(x+iy/2)/(x-iy/2)$.  $\eta_p$
and $\xi_p$ are phase factors related to the presence of a chemical potential
acting on the boundary.  Similarly, $\eta_{\mathrm{imp}}$ and
$\xi_{\mathrm{imp}}$ are phase factors associated to the impurity.  Their
explicit form will be given in the discussion of the different boundary and
impurity configurations below.  The phase shifts due to both scattering off
the boundary and the impurity generate corrections of order $L^0$ to
thermodynamical quantities.

We begin with a discussion fo the effect of the boundary potential: without
any boundary field, $p=0$, the ground state configuration of the open SUSY
$t$--$J$ chain is known to be given by real spin rapidities
$\{\lambda_j\}_{j=1,N_h +N_\uparrow}$ and hole rapidities
$\{\vartheta_\ell\}_{\ell=1,N_h}$ of the BAE (\ref{Sup}).  As the boundary
field is `switched on', however, purely imaginary solutions of the BAE become
possible and may have to be taken into account in the ground state.  Those
particular imaginary roots will be interpreted as boundary bound states (BBS)
induced by the local field (see also \cite{SkSa95,KaSk96,EsFr97}). Here we
shall distinguish three regimes depending on the value of the boundary
potential $p$:\newline
(i) For $0<p<1$, no BBS is solution to the BAE. The boundary phase factors are
the original ones derived by Essler \cite{Essl96}, i.e.\
\begin{equation}
  \eta_p(\lambda)\equiv 1\,,\quad
  \xi_p(\vartheta)=-e_{2/p-2}(\vartheta)\,.
\label{psbound}
\end{equation}
(ii) When $1 \le p <2$, the BAE allow for an imaginary solution for the hole
rapidities which we denote by $\vartheta_0 =i(1-{1 \over p})$ ($\mathrm{Im}\,
\vartheta_0\ge 0$).  Analysis of the spectrum implies that this root is
present in the ground state in the region $1 \le p <2$.  Taking the BBS
$\vartheta_0$ into account explicitely, the boundary phase factors in
(\ref{Sup}) become $\eta_p(\lambda)= e_{3-2/p}(\lambda) e_{2/p-1}(\lambda)$
and $\xi_p(\vartheta)= -e_{2/p-2}(\vartheta)$ with $N_h -1$ remaining real
roots $\vartheta_\ell$.
\newline
(iii) Increasing $p$ further, an additional BBS solution for a spin rapidity
arises in the thermodynamic limit: $\lambda_0 =i({1 \over 2}-{1 \over p})$
($\mathrm{Im}\, \lambda_0 \ge 0$).  Once again, both $\vartheta_0$ and
$\lambda_0$ have to be considered for the ground state and the effective
boundary phase factors for the remaining $N_h-1$ real hole rapidities
$\vartheta$ and $N_h +N_\downarrow -1$ real spin rapidities $\lambda$ become
$\eta_p(\lambda)=e_{-1-2/p}(\lambda)e_{-1+2/p}(\lambda)$ and
$\xi_p(\vartheta)=-e_{2/p}(\vartheta)$.
\newline
Note that for $p<0$ a complex solution $\vartheta_0$ always exists as the
condition $\mathrm{Im}\, \vartheta_0 \ge 0$ is trivially satisfied. This leads
to the same structure of boundary bound states as for positive $p$, i.e.\
repulsive boundary potential for hole excitations.  As a consequence this BBS
will not be part of the ground state configuration \cite{EsFr97}.  In the
following we restrict ourselves to strictly positive values of the boundary
potential.

A different, but completely equivalent description of the spectrum of the open
$t$--$J$ model with boundary impurity can be obtained by starting from the
Fock vacuum $|\Omega_L\rangle \equiv |0\rangle^{\otimes L}\otimes
|s-\frac{1}{2}\rangle_\mathrm{imp}$ (the so-called Lai pseudo-vacuum
\cite{Lai74}).  In this case the many-particle wave functions are parametrized
by $N_e$ charge rapidities $w_k$ and $N_\downarrow$ spin rapidities $x_\ell$
which solve the following set of BAE
\begin{eqnarray}
[e_1(w_k)]^{2L}\Phi_{\mathrm{imp}}(w_k)\Phi_p(w_k) =
	\prod_{\ell=1}^{N_\downarrow}e_1(w_k - x_\ell)
	e_1(w_k + x_\ell)\,,
\nonumber\\[-8pt]
\label{Lap}
\\[-8pt]
\nonumber
  \Xi_{\mathrm{imp}}(x_\ell)\Xi_p(x_\ell)
     \prod_{j=1}^{N_e} e_1(x_\ell - w_j) e_1(x_\ell
  + w_j) = \prod_{m\ne \ell}^{N_\downarrow}e_2(x_\ell - x_m) e_2(x_\ell +
  x_m)\,. 
\end{eqnarray}
Just as in the Sutherland equations (\ref{Sup}), $\Phi_p$ and $\Xi_p$
(resp. $\Phi_{imp}$ and $\Xi_{imp}$) are the phase factors associated to the
boundary potential (resp. the impurity).  The equivalence between the Lai and
the Sutherland description of the model can proved on the basis of a
particle--hole ($p$--$h$) transformation at the level of the BAE
\cite{Woyn83,BarX92,EsKo92} (see \ref{Apph}).  As an immediate consequence of
this $p$--$h$ symmetry the boundary phases $\eta_p$, $\xi_p$ in the different
$p$-regimes identified above can be mapped to $p$-dependent phases in  the Lai
formulation (\ref{Lap}) of the BAE:
(i) Starting from the `bare' Sutherland equations (\ref{Sup}) with
(\ref{psbound}) (i.e.\ without occupied BBS)  the $p$--$h$ transformation
(\ref{Map}) gives 
\begin{equation}
  \Phi_p(w)=-e_{2/p-1}(w)\,,\quad \Xi_p(x)\equiv 1\,.
\label{plbound}
\end{equation}
The analysis of of the Lai BAE (\ref{Lap}) with these boundary phases shows
that no BBS exist in this formulation for $p<2$.
\newline 
(ii) For $p\ge 2$, $w_0=i({1 \over 2}-{1 \over p})$ is solution to the BAE
(with $\mathrm{Im}\, w_0 \ge 0$). It corresponds to a charge-like bound state
in the Lai sector which increases the energy of the state.  Therefore it is
not part of the ground state configuration which is still described by
Eqs.~(\ref{Lap}) with the boundary phases given before.  Populating the bound
state $w_0$, one obtaines a different part of the spectrum which is described
by (\ref{Lap}) with modified boundary phases $\Phi_p(w)=-e_{2/p-1}(w)$ and
$\Xi_p(x)=e_{2/p}(x)e_{2-2/p}(x)$.  Furthermore, since the occupation of the
BBS is taken into account explicitely, the number of charge rapidities has to
be lowered by one, $N_e \to N_e -1$.

A similar sequence of bound states appear when the coupling of the impurity is
varied by changing the parameter $t$ in Eq.~(\ref{kdress}).  Starting from the
state with maximal polarization $|\Omega_S\rangle=|\uparrow\rangle^{\otimes
L}\otimes|s\rangle$ the spectrum is determined by BAE of Sutherland type
(\ref{Sup}) with
\begin{equation}
 \eta_{\mathrm{imp}}(\lambda) = e_{2s}(\lambda -t)e_{2s}(\lambda +t)\,,
 \quad \xi_{\mathrm{imp}}(\vartheta)\equiv 1\,,
\label{psimp}
\end{equation}
while in the corresponding phase shifts in the Lai formulation of the BAE read
\begin{eqnarray}
    \Phi_\mathrm{imp}(w)=e_{2s}(w_k +t)e_{2s}(w_k -t)\,,\nonumber\\
    \Xi_\mathrm{imp}(x)=e_{2s-1}(x_\ell +t)e_{2s-1}(x_\ell -t)\,.
\label{plimp}
\end{eqnarray}

The additional phases, e.g.\ $\eta_{\mathrm{imp}}$ from (\ref{psimp}) in
(\ref{Sup}), allows for new imaginary solutions to the BAE which can be
interpreted as impurity bound states (IBS) similar as in a continuum model
related to the Kondo problem \cite{GoHs90,WaVo96} and for an Anderson-type
impurity in the $t$--$J$ model \cite{FrPa06}.  They appear for $t$ being a pure
imaginary number itself, $t=i\tau$ with $\tau\in\mathbbm{R}^+$.  A short
analysis of the Eqs. (\ref{Sup}) with (\ref{psimp}) in the thermodynamic limit
reveals that there are two absolute thresholds opening an IBS:\newline (i) If
$\tau \ge s$, $\lambda_0 = i(\tau-s)$ is a IBS solution with $\mathrm{Im}\,
\lambda_0 \ge 0$.\newline (ii) For $\tau \ge s+1/2$, a $\vartheta$-IBS
appears, $\vartheta_0 = i(\tau -s-1/2)$ ($\mathrm{Im}\, \vartheta_0 \ge 0$).

\section{The SUSY $t$--$J$ model with a boundary Kondo spin} %
\label{sec:kondo}
Up to now, the effect of the boundary potential $p$ and the presence of the
impurity on the spectrum of the $t$--$J$ chain has been discussed separately.
This approach covers the generic case of an impurity described by a $c$-number
solution to the reflection equations (\ref{REF}) dressed by an ${\cal
L}$-operator which describes the inhomogeneity (\ref{kdress}).  Such solutions
to the RE are called `regular' in opposition to other `singular' solutions
which cannot be obtained by the dressing prescription.  For the issue of
impurities in the $t$--$J$ model, `singular' boundary matrices have been
obtained by Zhou \emph{et al.} \cite{ZhGo99,ZGLG99}.  As has been shown in
Ref.~\cite{FrSl99} these singular boundary matrices can be obtained from
regular ones by suitable adjusting the parameters describing the impurity ($s$
and $t$ in our case) and the boundary potential $p$ followed by a projection
onto a subspace of the impurity Hilbert space ${\cal H}$.
To apply this `projecting method' ${\cal H}$ is decomposed into two orthogonal
subspaces ${\cal H}_1$ and ${\cal H}_2$, such that ${\cal H}_1 \oplus{\cal
H}_2={\cal H}$.  To each of these subspaces we associate the projectors
$\Pi_1$ and $\Pi_2$. Then a necessary and sufficient condition for the
projections $\Pi_i {\cal K}_-(\lambda)\Pi_i$ of (\ref{kdress}) on these
subspaces to satisfy the RE is the vanishing of one of the projections
\begin{equation}
  \Pi_1 {\cal K}_-(\lambda)\Pi_2=0,\qquad
  \mbox{or}\qquad \Pi_2 {\cal K}_-(\lambda)\Pi_1=0.
\end{equation}
The `projected' boundary matrices resulting from this construction are
`singular' as shown in Ref.~\cite{FrSl99}.

Here we apply the projection to the inhomogeneity described by the ${\cal
L}$-operator (\ref{LS+}).  A natural decomposition of the impurity's quantum
space is onto the subspaces spanned by the two different spin multiplets
contained in $[s]_+$ (see \ref{app:gl21}), namely ${\cal H}_1 =
\mathrm{span}\{|s, s,m\rangle\}$ and ${\cal H}_2 =
\mathrm{span}\{|s+\frac{1}{2}, s-\frac{1}{2},m\rangle\}$.  With this
decomposition one findes that $\Pi_1 [{\cal
L}_\mathrm{imp}(\lambda+t)K_-^p(\lambda)\left({\cal
L}_\mathrm{imp}(-\lambda+t)\right)^{-1}] \Pi_2$ vanishes for
\begin{equation}
 {t}=i\left(-{1 \over p} - s + {1 \over 2}\right)\equiv i \tilde{t}.
\label{tproj}
\end{equation}
while $\Pi_2 {\cal K}_-(\lambda) \Pi_1$ vanishes for $t=-i\tilde{t}$. Both
projections are actually equivalent and give rise to the same effective
Hamiltonians within the two spin subsectors.

As a consequence of the projection the boundary Hamiltonian (\ref{Hbimp}) of
the impurity system simplifies giving a purely magnetic impurity of spin $s$
or $s-\frac{1}{2}$.  Hence the model is that of a Kondo impurity coupled to
the edge of a one-dimensional model of correlated electrons.
For the special case $s=\frac{1}{2}$ the projection $\Pi_1$ gives the simple
Hamiltonian studied by Wang \emph{et al.} \cite{WDHP97,DW98}):
\begin{equation}
H_\mathrm{bimp}|_{s=1/2}=J_0\, \mathbf{S}_1\cdot\mathbf{s} +V_0\, n_1
\end{equation} 
where $\mathbf{s}$ is the impurity spin operator and $J_0 = 2p^2 /(p^2 -4)$
within our notations.  The fine-tuning of boundary and impurity terms
necessary to ensure integrability of the Hamiltonian (\ref{Hbimp}) in
combination with the projection leads to the pure Kondo exchange with a
coupling constant $J_0$ controlled by the remaining free parameter, i.e.\ the
boundary potential $p$.
  
\begin{figure}[ht]
\begin{center}
\includegraphics[width=0.5\textwidth]{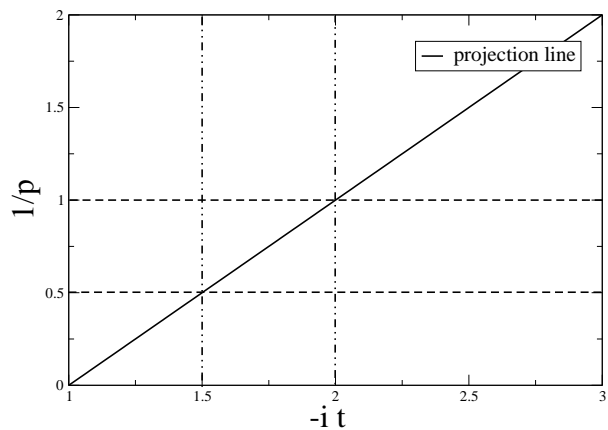}
\caption{Intersection of the IBS (dash-dotted lines) and BBS
  thresholds (dashed lines) with the `projection line' (solid
  line) (\ref{tproj}) for $s=3/2$ as an example.}
\label{projFig}
\end{center} 
\end{figure}

In the remainder of this paper we study the question, how the spectrum of the
projected system emerges from the original one, i.e.\ how the BAE introduced
above have to be modified for the projected Kondo-type Hamiltonian.
After synchronizing the impurity and boundary parameters to the `projecting
line' Eq. (\ref{tproj}), the sequences of BBS and IBS are no longer
independent.  Instead, one finds, that IBS' thresholds now coincide with the
BBS ones exactly, leaving only one sequence of BS to take care of (see
Fig. \ref{projFig}).

We begin by enforcing the projection condition (\ref{tproj}) in the Sutherland
equations (\ref{Sup}) for the small $p$ regime (\ref{psbound}) with
(\ref{psimp}).  This results in:
\begin{eqnarray}
  [e_1(\lambda_k)]^{2L}e_{1-2/p}(\lambda_k)e_{4s+2/p-1}(\lambda_k) =
  \prod_{j\ne k}^{N_h
    +N_\downarrow}e_2(\lambda_k-\lambda_j)e_2(\lambda_k+\lambda_j)
  \times
\nonumber\\
  \qquad\qquad \times\prod_{\ell=1}^{N_h}e_{-1}(\lambda_k
  -\vartheta_\ell)e_{-1}(\lambda_k+\vartheta_\ell)\,,
\label{SuSP}\\
  1= -e_{2/p-2}(\vartheta_\ell)\prod_{j=1}^{N_h
    +N_\downarrow}e_1(\vartheta_\ell -\lambda_j)e_1(\vartheta_\ell
    +\lambda_j)\,.
\nonumber
\end{eqnarray}
Since the Sutherland BA starts from the fully polarized state, in particular
the state $|s\rangle_\mathrm{imp}$ for the impurity site, the solution to
these equations will describe the spectrum of (\ref{HtJ}), (\ref{Hbimp})
restricted to the spin-$s$ subspace ${\cal H}_1$, i.e. that of a Kondo spin
$s$ impurity in a correlated $t$--$J$ chain.

Alternatively, we can study the other sector of the impurity Hilbert space
selected by the projection scheme, namely ${\cal H}_2$ which is the impurity
spin $s-1/2$ subspace.  In this case, however, the BAE have to be derived from
a different pseudo vacuum since the initial state $|s\rangle$ of the impurity
used in the Sutherland approach will be discarded by the projection
prescription.  A possibility to circumvent this problem is to implement the
projection condition (\ref{tproj}) to the Lai BAE (\ref{Lap}) with
(\ref{plbound}), (\ref{plimp}) directly.  Here the pseudo vacuum used in the
algebraic Bethe ansatz is an element of the projected subspace.  Hence, the
spectrum of the $\Pi_2$ projected impurity is determined by the roots of the
Lai-type BAE
\begin{eqnarray}
   -[e_1(w_k)]^{2L}e_{4s+2/p-1}(w_k) =
  \prod_{\ell=1}^{N_\downarrow}e_1(w_k - x_\ell)e_1(w_k + x_\ell)\,,
\nonumber\\
  \prod_{j=1}^{N_e} e_1(x_\ell - w_j)e_1(x_\ell + w_j) =
  e_{2/p}(x_\ell)e_{2-4s-2/p}(x_\ell)\times
\label{LaSP}\\
    \qquad\qquad
  \times\prod_{m \ne \ell}^{N_\downarrow}e_2(x_\ell - x_m)
    e_2(x_\ell + x_m)\,.
\nonumber
\end{eqnarray}

Up to this point we have made repeated use of the fact that the BAE for the
unprojected Hamiltonian in the Sutherland and Lai formulation are related by
the $p$--$h$ transformation described in \ref{Apph}.  Now let us apply the
$p$--$h$ transformation to the projected equations to see whether and how this
relation manifests itself in the latter.  Using Eqs.~(\ref{Map}) on the
projected Sutherland BAE (\ref{SuSP}) we obtain the following system of
equations in the Lai sector:
\begin{eqnarray}
  -[e_1(w_k)]^{2L}e_{4s+2/p-1}(w_k) = \prod_{\ell=1}^{N_\downarrow}
     e_1(w_k - x_\ell)e_1(w_k + x_\ell)\,,
\nonumber\\
\label{LaSPBS}
  e_{2-2/p}(x_\ell)\prod_{j=1}^{N_e -1} e_1(x_\ell - w_j) e_1(x_\ell + w_j)
  = \\
   \qquad\qquad =  e_{2-4s-2/p}(x_\ell)\,
  \prod_{m \ne \ell}^{N_\downarrow} e_2(x_\ell - x_m)
  e_2(x_\ell + x_m)\,.
\nonumber
\end{eqnarray}
As expected from our discussion above, these differ from the Lai projected BAE
(\ref{LaSP}).  Instead Eqs. (\ref{LaSPBS}) are the BAE that one would obtain
after populating the charge bound state $w_0$ explicitly.  Both sets of BAE
(\ref{SuSP}) and (\ref{LaSPBS}) can be used for studying the spin-$s$ subspace
${\cal H}_1$ of the impurity.  The problem of working with a proper
pseudo-vacuum in the Lai sector is overcome by enforcing the occupation of the
bound state.  In fact, Eqs. (\ref{LaSPBS}) coincide with the rational limit of
the Bethe equations for the $q$-deformed supersymmetric $t$--$J$ model with
spin impurities described in terms of singular boundary matrices, i.e.\
working directly in the projected impurity Hilbert space
\cite{FaWY00,FaWa01,GeGZ01}.

Similarly, using Eqs. (\ref{Map}), to map the projected Lai BAE (\ref{LaSP})
giving the spectrum in the ${\cal H}_2$ sector one obtains $p$--$h$
transformed equations for this sector in the Sutherland formulation:
\begin{eqnarray}
  [e_1(\lambda_k)]^{2L}e_{-1-2/p}(\lambda_k)e_{4s+2/p-1}(\lambda_k) =
  \prod_{j\ne k}^{N_h
    +N_\downarrow}e_2(\lambda_k-\lambda_j)e_2(\lambda_k+\lambda_j)\times
\nonumber\\
  \qquad\qquad\times\prod_{\ell=1}^{N_h}e_{-1}(\lambda_k
  -\vartheta_\ell)e_{-1}(\lambda_k+\vartheta_\ell)\,,
\label{SuSPBS}
\\
  1= -e_{2/p}(\vartheta_\ell)\prod_{j=1}^{N_h
    +N_\downarrow}e_1(\vartheta_\ell -\lambda_j)e_1(\vartheta_\ell
    +\lambda_j)\,.
\nonumber
\end{eqnarray}
Comparing the boundary phase shifts in these equations we identify those BAE
with the ones obtained by occupying the $\lambda_0$- and $\vartheta_0$-BBS
discussed in Section \ref{sec:bbs}.
%

\section{Summary}
We have analyzed the role of the bound states appearing in the Bethe ansatz
solution for a lattice system with open boundaries.  These particular
solutions indeed correspond to localized objects at the site of the impurity.
As the boundary chemical potential $p$ and the spectral shift $t$ are
synchronized according to (\ref{tproj}) charge fluctuations are suppressed
completely which allows for the projection onto the Hilbert space of a purely
magnetic impurity.  The equivalence of the Bethe ans\"atze obtained from
different reference states relies on the proper choice of the bound state
configurations to select the relevant sector of states as has already been
conjectured in \cite{BeFr99a}.

Finally, let us note that the Sutherland equations (\ref{SuSP}),
(\ref{SuSPBS}) and similarly the Lai equations (\ref{LaSP}), (\ref{LaSPBS})
for the different impurity sectors of the model are mapped into each other by
means of the replacement $1/p\to 1/p+1$ and $s\to s-\frac{1}{2}$.  This
follows from the fact that the same impurity model can be obtained by
projection onto the spin $s-\frac{1}{2}$ sector of either an $[s]_+$- or an
$[s-\frac{1}{2}]_+$-inhomogeneity.  Note that the synchronization condition
(\ref{tproj}) has to be adapted accordingly.
To gain additional insight into the underlying algebraic structure of the
projection mechanism and the relation between the resulting spin $s$- and
$s-\frac{1}{2}$-sectors as well as into the thermodynamical properties of a
Kondo-spin coupled to a correlated host the Bethe equations in Section
\ref{sec:kondo} have to be analyzed in more detail.  This will be the subject
of a forthcoming study.


\ack
This work has been supported by the Deutsche Forschungsgemeinschaft.

\appendix
\section{The (super)algebra $gl(2|1)$}  %
\label{app:gl21}
Apart from the generators $1$, $S^z$, $S^{\pm}$ forming an (ungraded) $gl(2)$
subalgebra, $gl(2|1)$ has an additional generator $B$ of even parity (charge),
commuting with the spin operators, and four odd parity generators $V^\pm$ and
$W^\pm$.  The commutation relations between even and odd generators are listed
below:
\begin{equation}
\begin{array}{c}
[S^z,V^\pm]=\pm\frac{1}{2} V^\pm,\qquad [S^\pm,V^\pm]=0,\qquad
[S^\mp,V^\pm]=V^\mp,\nonumber\\
{[S^z,W^\pm]}=\pm\frac{1}{2} W^\pm,\qquad [S^\pm,W^\pm]=0,\qquad
[S^\mp,W^\pm]=W^\mp,\nonumber\\
{[B,V_\pm]}=\frac{1}{2} V_\pm,\qquad [B,W_\pm]=-\frac{1}{2} W_\pm.
\end{array}
\end{equation}
The odd generators satisfy anticommutation relations
\begin{equation}
\begin{array}{c}
\{V^\pm,V^\pm\}=\{V^\pm,V^\mp\}=
\{W^\pm,W^\pm\}=\{V^\pm,W^\mp\}=0,\nonumber\\
\{V^\pm,W^\pm\}=\pm\frac{1}{2} S^\pm,\qquad
\{V^\pm,W^\mp\}=\frac{1}{2} (S^z\pm B).
\end{array}
\end{equation}
The irreducible representations of $gl(2|1)$ can be classified into typical
and atypical ones \cite{snr:77,Marcu80}.  With respect to to the even parity
$U(1)$ and $SU(2)$ subalgebras they can be decomposed into spin multiplets and
are conveniently labelled by the eigenvalues of the even parity operators $B$,
$\mathbf{S}^2$ and $S^z$.  The typical $8s$-dimensional representation $[b,s]$
contains four spin-multiplets 
\begin{equation*}
\begin{array}{l}
  \{|b,s,m\rangle,\, m=-s,\ldots,s\}\,,\\
  \{|b,s-1,m\rangle,\, m=-s+1,\ldots,s-1\}\,,\\
  \{|b\pm\frac{1}{2},s-\frac{1}{2},m\rangle,\,
    m=-s+\frac{1}{2},\ldots,s-\frac{1}{2}\}\,.
\end{array}
\end{equation*}
As $b\to\pm s$ these representations degenerate into two atypical ones.
Atypical representations are denoted by $[s]_\pm$ and contain $4s+1$ states in
two spin multiplets
\begin{equation*}
\begin{array}{l}
  \{|\pm s,s,m\rangle,\,m=-s,\ldots,s\}\,,\\
  \{|\pm (s+\frac{1}{2}),s-\frac{1}{2},m\rangle,\,
      m=-s+\frac{1}{2}),\ldots,s-\frac{1}{2}\}
\end{array}
\end{equation*}
respectively.  When no confusion is possible we denote the $SU(2)$ highest
weight states in the atypical representation $[s]_+$ by $|s,s,s\rangle \equiv
|s\rangle$ and $|s+\frac{1}{2},s-\frac{1}{2},s-\frac{1}{2}\rangle \equiv
|s-\frac{1}{2}\rangle$.

The superalgebra $gl(2|1)$ has two Casimir operators, we have used the
quadratic one
\begin{equation}
\label{cas2}
C_2 = B^2-\mathbf{S}^2 + W_- V_+ - W_+ V_- + V_- W_+ - V_+ W_-
\end{equation}
to express the ${\cal L}$-operator and the Hamiltonian in the main text.  On a
typical representation $[b,s]$, $C_2$ takes the value $b^2-s^2$ while it
vanishes on the atypical ones for any $s$.
%
%
%

\section{$p$\ --\ $h$ transformation of the BAE: Lai and Sutherland pseudo
  vacuua}
\label{Apph}
There are three different BAE for the $gl(2|1)$ supersymmetric $t$--$J$ model
depending on the choice of grading in the algebra that contains two fermions
and one boson \cite{EsKo92}.  Here we will focus on two equivalent
constructions of the spectrum of the $t$--$J$ model with an $[s]_+$-impurity
which differ in the choice of the highest-weight state used for the pseudo
vacuum in the algebraic Bethe ansatz.  Either one can construct the Bethe
states starting from the so-called Lai vacuum $|\Omega_L\rangle =
|0\rangle^{\otimes L} \otimes
|s+\frac{1}{2},s-\frac{1}{2},s-\frac{1}{2}\rangle_{imp}$ or from the so-called
Sutherland vacuum $|\Omega_S \rangle = |\uparrow\rangle^{\otimes L} \otimes
|s,s,s\rangle_{imp}$.  The two approaches are perfectly equivalent for the
description of the system's spectrum. In the case of the homogeneous chain
with periodic boundary conditions, the equivalence of the Lai and Sutherland
BAE has been proven in Ref. \cite{BarX92} using a $p$--$h$ transformation
introduced by Woynarovich \cite{Woyn83}.  The aim of this Appendix is to
generalize this technique to open boundary conditions including the
possibility of having boundary fields and impurity phase shifts.  Nevertheless
the spirit of the proof is very similar to the one derived in the periodic
case.

\begin{figure}[ht]
\begin{center}
\includegraphics[width=0.6\textwidth]{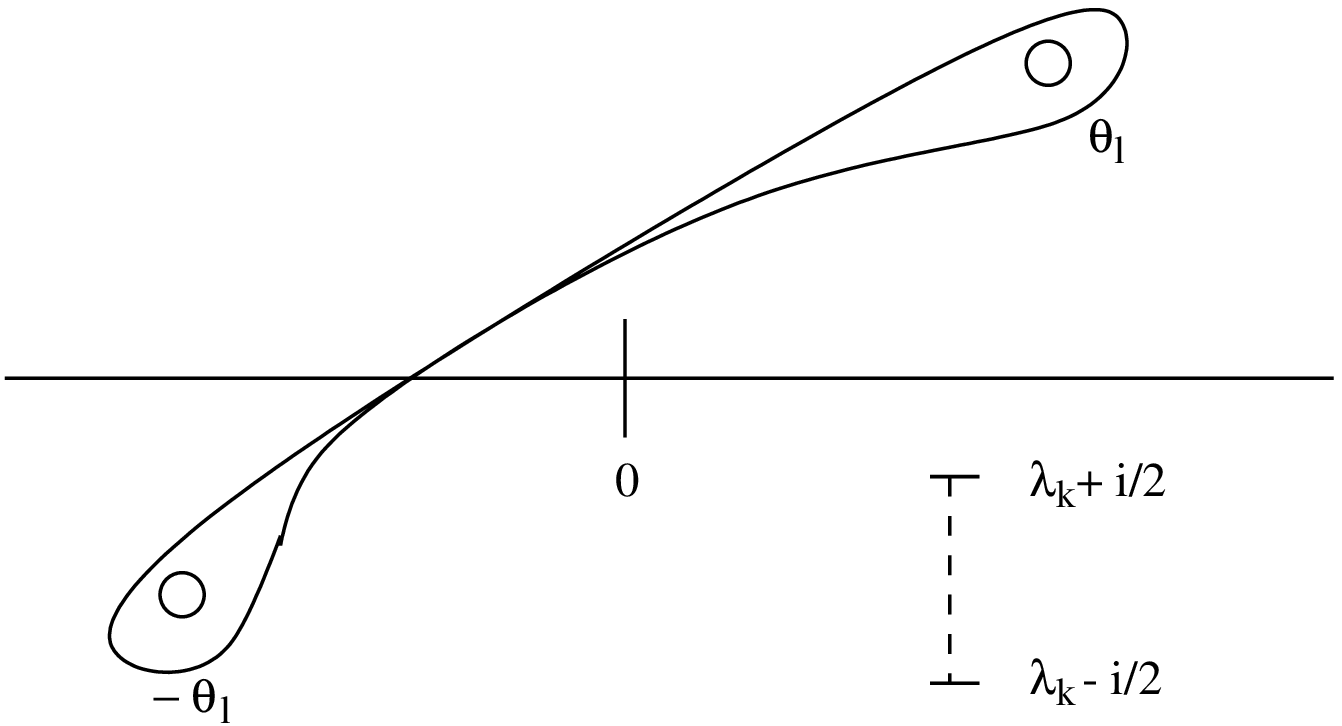}
\caption{The Contour ${\cal C}_\ell$ used in Eq. (\ref{defC}) encloses the
  points $\vartheta_\ell$ and $-\vartheta_\ell$. The branch cut of the
  logarithm in the integrand is depicted as the dashed line connecting from
  $\lambda_k -i/2$ to $\lambda_k+i/2$.}
\label{FigCl}
\end{center}
\end{figure}

If one starts from the Sutherland vacuum, the resulting BAE for a $t$--$J$
model with boundaries are given by:
\begin{eqnarray}
\fl
  [e_1(\lambda_k)]^{2L}\eta(\lambda_k) &=
  \prod_{j\ne k}^{N_h
    +N_\downarrow}e_2(\lambda_k-\lambda_j)e_2(\lambda_k+\lambda_j)
  \prod_{\ell=1}^{N_h}e_{-1}(\lambda_k-\vartheta_\ell)
  e_{-1}(\lambda_k+\vartheta_\ell)\,,
\nonumber\\[-8pt]
\label{Sub}\\[-8pt]
  1&= \xi(\vartheta_\ell)\prod_{j=1}^{N_h
    +N_\downarrow}e_1(\vartheta_\ell -\lambda_j)e_1(\vartheta_\ell
  +\lambda_j)\,.
\nonumber
\end{eqnarray}
where $\eta$ and $\xi$ are phase factors (rational functions in their
arguments) describing the boundary and inhomogeneity scattering (see
Eqs.~(\ref{Sup})).
>From the second set of these equations we find that $\vartheta_\ell$ are
zeroes of the polynomial
\begin{eqnarray}
P(w)&=&\xi_+ (w) \prod_{j=1}^{N_h +N_\downarrow} (w-\lambda_j +{i \over
2}) (w +\lambda_j +{i \over 2})\nonumber\\
&&- \xi_- (w)\prod_{j=1}^{N_h
+N_\downarrow} (w-\lambda_j -{i \over 2}) (w +\lambda_j -{i \over 2}) \equiv 0.
\label{Pw}
\end{eqnarray}
Here $\xi_+$ (resp. $\xi_-$) stands for the numerator (resp. denominator) of
the function $\xi$.  $P(w)$ is of degree $2(N_h + N_\downarrow) + \delta$
where $\delta$ is determined by the degree and the parity of $\xi_\pm
(w)$. Hence, in addition to the first $2N_h$ roots of $P(w)$ which we
identify with the roots $\{\vartheta_\ell\}$ of the BAE (\ref{Sub}) there are
$2N_\downarrow + \delta$ additional zeroes $\{x_\ell\}$. Notice
that $P(w)$ is an odd polynomial in all cases considered in this paper.
Consequently, the zeroes of $P$ come in pairs $\vartheta_\ell =
-\vartheta_{-\ell}$ and $x_\ell = -x_{-\ell}$ except from a single root at
$x_0 = 0$.  Using the residue theorem we obtain:
\begin{eqnarray}
\fl
\sum_{\ell=1}^{N_h} {1 \over i}\ln\left(\frac{\lambda_k
  -\vartheta_\ell -{i 
\over 2}} {\lambda_k -\vartheta_\ell +{i \over 2}}\frac{\lambda_k
+\vartheta_\ell -{i \over 2}} {\lambda_k +\vartheta_\ell +{i \over
2}}\right) =
\sum_{\ell=1}^{N_h} \frac{1}{2\pi i}\oint_{{\cal C}_\ell} \dd
z {1 \over i} \ln\left( \frac{\lambda_k -z -{i \over 2}}{\lambda_k -z
+{i\over 2}}\right) \frac{\dd}{\dd z}\ln P(z)=
\nonumber\\[-8pt]
\label{defC}
\\[-8pt]
\fl
=-\sum_{\ell=1}^{N_\downarrow} {1 \over i} \ln\left(
\frac{\lambda_k - x_\ell -{i \over 2}}{\lambda_k - x_\ell +{i\over
2}}\,\frac{\lambda_k + x_\ell -{i\over 2}} {\lambda_k + x_\ell +{i\over
2}}\right)
 - {1 \over i}\ln\left(\frac{\lambda_k -
{i\over 2}} {\lambda_k + {i\over 2}}\right) + {1\over i} \ln
\left(\frac{P(\lambda_k -{i\over 2})} {P(\lambda_k + {i\over
2})}\right)
\nonumber
\end{eqnarray}
(the last sum runs over the nonzero $x_\ell$).
The contour ${\cal C}_\ell$ is chosen such that it encloses both zeroes
$\vartheta_\ell$ and $-\vartheta_\ell$ carefully avoiding the logarithm's
branch cut between $\lambda_k -i/2$ and $\lambda_k+i/2$ (see
Fig.~\ref{FigCl}).
By definition of $P$ (\ref{Pw}) we evaluate its value at both ends of the
branch cuts,
\begin{eqnarray}
P(\lambda_k - {i \over 2}) &=& -\xi_- (\lambda_k - {i\over
2})\prod_{j=1}^ {N_h + N_\downarrow}(\lambda_k -\lambda_j
-i)(\lambda_k +\lambda_j -i)\,,
\nonumber\\[-8pt]
\label{Pbranch}\\[-8pt]
\nonumber
P(\lambda_k + {i \over 2}) &=& \xi_+
(\lambda_k + {i\over 2})\prod_{j=1}^ {N_h + N_\downarrow}(\lambda_k
-\lambda_j +i)(\lambda_k +\lambda_j +i)\,.
\end{eqnarray}
Exponentiating Eq. (\ref{defC}) we obtain
\begin{eqnarray}
&&\prod_{\ell=1}^{N_h} e_{-1}(\lambda_k - \vartheta_\ell)e_{-1}(\lambda_k
+\vartheta_\ell)
=  -e_1 (\lambda_k)\,\frac{\xi_- (\lambda_k - {i \over
2})} {\xi_+ (\lambda_k + {i \over 2})}\times 
\nonumber\\ 
&&\quad\times
\prod_{\ell=1}^{N_\downarrow} e_1(\lambda_k-x_\ell) e_1(\lambda_k + x_\ell)\,
\prod_{j=1}^{N_h +N_\downarrow} e_{-2}(\lambda_k -\lambda_j)
e_{-2}(\lambda_k +\lambda_j)\,.
\label{expD}
\end{eqnarray}
The last product appearing on the r.h.s. can be reexpressed as
\begin{equation}
\fl
\prod_{j=1}^{N_h +N_\downarrow}e_{-2}(\lambda_k -\lambda_j)
e_{-2}(\lambda_k +\lambda_j) = -e_{-1}(\lambda_k)\prod_{j\ne k}^{N_h
  +N_\downarrow} e_{-2}(\lambda_k -\lambda_j) e_{-2}(\lambda_k +\lambda_j)
\end{equation}
since $e_{-2}(2\lambda_k)=e_{-1}(\lambda_k)$.  Then, using Eq. (\ref{expD}) in
the first of Eqs.~(\ref{Sub}) we obtain:
\begin{equation}
\eta(\lambda_k)[e_1 (\lambda_k)]^{2L} =  \frac{\xi_- (\lambda_k
-{i\over 2})} {\xi_+ (\lambda_k +{i\over 2})}\,
\prod_{\ell=1}^{N_\downarrow}e_1
(\lambda_k -x_\ell) e_1 (\lambda_k + x_\ell)\,
\label{Lab1}
\end{equation}
(cf.\ the first of the Lai equations (\ref{Lap})). Starting from
Eq. (\ref{Lab1}), it is straightforward to apply the same procedure as before
to derive the second Lai type equation.

To summarize the main result of this Appendix let us write the relation
connecting the boundary phase factors within BAE in the Sutherland
representation (\ref{Sub}) and those in the Lai representation:
\begin{eqnarray}
&& \Phi(w_k)[e_1 (w_k)]^{2L} = 
  \prod_{\ell=1}^{N_\downarrow}e_1 (w_k -x_\ell) e_1 (w_k +x_\ell),
\nonumber\\[-8pt]
\label{LaiBAE}\\[-8pt]
\nonumber
&& \Xi(x_\ell)\prod_{j=1}^{N_e}e_1 (x_\ell -w_j) e_1 (x_\ell +w_j) 
  = \prod_{m\ne\ell}^{N_\downarrow} e_2 (x_\ell - x_m) e_2 (x_\ell +x_m).
\end{eqnarray}
Comparing the result of the particle-hole transformation applied to
(\ref{Sub}) with (\ref{LaiBAE}) we find
\begin{equation}
\eta(\lambda)\frac{\xi_+ (\lambda +{i\over 2})}{\xi_- (\lambda
  -{i\over 2})} = \Phi(\lambda)\,,
\qquad
\xi^{-1}(x) = \Xi (x) \frac{\Phi_- (x + {i\over 2})}{\Phi_+ (x
-{i \over 2})}\,.
\label{Map}
\end{equation}


\section*{References}

\end{document}